# A Strain-Engineered 0D/1D Heterojunction of $InVO_4$/Cu-$TbFeO_3$ for High-Selectivity $CO_2$ Photoreduction

Muneeb Ur Rahman[1]

[1]Collaborative Innovation Center of Biomass Energy, Henan Agricultural University, Zhengzhou 450002, China

**Abstract**

The catalytic $CO_2$ photoreduction to CO is significantly hindered by the pervasive kinetic bottleneck of *CO-desorption and inefficient charge separation. Surpassing the conventional single photocatalytic strategy, herein, a multi-synergetic 0D/1D S-scheme heterojunction by precisely assembling 0D $InVO_4$ nanoparticles on 1D Cu-doped $TbFeO_3$ (IVO/CTFO). This nano-heterojunction is rationally designed at multiple steps where $Cu^{2+}$ substitution at the $Fe^{3+}$ site induces a compression in lattice strain and oxygen vacancies ($V_O$), acting as electron traps and $CO_2$ chemisorption sites, which breaks spin-polarization of pristine $TbFeO_3$ to facilitate multichannel charge flow. The 0D/1D strategy couples the maximum surface active-sites and short charge diffusion routes with directional charge migration. Moreover, a 0D/1D lattice mismatch creates a built-in electric field at the interface, resulting in an enhanced lifetime (64.70 ns) of charged species, an efficient CO yield (65.75 µmole $g^{-1}.h^{-1}$), and high selectivity (95.93%). DFT calculations and experimental findings confirmed the Fermi level shift toward the conduction band and the existence of spin-hybridization. Operando-DRIFTS and the free-energy diagram unveil a $H^+$ mediated mechanism at the interface, alongside a reduction in energy barrier for $CO_2$ photoreduction from *COOH to *CO. Thus, this study presents an excellent approach that integrates defect-engineering, strain-compression, and interfacial design in advancing solar fuels production.

**Introduction**

The constant accumulation of atmospheric $CO_2$ is surpassing 420 ppm, posing a serious risk to the environment and resulting in ocean acidification and climate change effects. The $CO_2$ photoreduction, a natural photosynthesis, is considered an excellent approach for the production of renewable fuel by mitigating secondary emissions. In particular, photoreduction of $CO_2$ possesses a significant value, where CO acts as a precursor for synthetic fuel production [1] in the Fischer-Tropsch reaction [2]. Additionally, C═O is stable with a bonding energy of ≈750 kJ/mol [3], and its kinetic barrier to multi-electron transfer has inhibited its practical applications. To breach this kinetic barrier, an advanced catalyst is required for the photoreduction of $CO_2$ to CO. Also, the adsorption along with activation of $CO_2$ on the surface of the catalyst are leading factor in the $CO_2$ photoreduction reaction. Addition of heteroatom elements into the catalyst lattice can effectively regulate the catalytic active sites along with their electronic structure, thereby enhancing the adsorption and activation of $CO_2$ and facilitating selective sorption of reaction intermediates [4]. As an excellent approach, the rational fabrication of semiconducting heterojunctions has garnered substantial attention, as the heterojunctions efficiently improve catalytic activity via $e^-h^+$ pairs separation and synergize the redox potentials of comprising materials, while their interfacial electric field dynamically suppresses the recombination rate of charged carriers [5]. Currently, several metal dopants (e.g., Mn, Ni, Cu, Ag, etc.) have been explored to increase $CO_2$ reduction performance [6] because of their flexible electronic framework, which improves their optical and electrochemical behavior with their potential to alter the characteristics of the host constituents, however some. In spite of this, Cu-substitution into the catalyst lattice can modify the electronic structure by introducing intermediate energy levels, it can enhance light-capturing ability by narrowing the band gap, and form additional active sites. Moreover, the conducive nature of Cu is responsible for $CO_2$-sorption, which yields $C_{2+}$ products [7]. Explicitly, Cu has been extensively studied as the metal demonstrating negative $^*CO$ adsorption energy, which can produce $C_{2+}$ products from $CO_2$ due to its optimum binding affinity for $CO_2$ and reaction intermediates. To increase the $CO_2$ photoreduction activity of catalysts and their stability, numerous regulating strategies have been proposed, including doping [8], formation of cation/anion vacancy [9], single atom engineering [10], and development of heterojunction [11]. Despite these advancements, the comprehensive reaction mechanism remains uncertain, underscoring the necessity for systematic analysis at the atomic level and the rational strategy in designing highly selective photocatalysts. Feng Yanmei et al. reported ligand-free Cu-doped ultrathin $Cs_3Bi_2Br_9$ nanoplates for $CO_2$ photoreduction to CO [12]. Li et al. worked on a type-II heterojunction with a core-satellite structure for photocatalytic $CO_2$ reduction [13]. Pan et al. prepared a type-II heterojunction g-$C_3N_4$/$Cs_2AgBiBr_6$ for photocatalytic $CO_2$ reduction [14]. In 2020, Yu et al. reported, for the first time, the application of $TiO_2$/$CsPbBr_3$ with S-scheme heterojunction for $CO_2$ photoreduction [15]. Xu et al. also prepared a $CuInS_2$/PCN S-scheme heterojunction with a close contact interface, which enhanced the $CO_2$ reduction efficiency [16].

Summarizing the above research findings reveals that the currently reported systems rely on a single strategy, either doping or heterojunction formation, and lack a systematic theory and method for their formation. Most of them are simply formed by selectively coupling two or more semiconductors, which suffer from low yields due to rapid recombination, dependence on sacrificial agents, and low selectivity. Therefore, it is imperative to rationally modulate photocatalyst structure to achieve higher yields, activity, and selectivity for practical $CO_2$ photoreduction under diluted conditions.

Herein, we bridge this gap by developing a 0D/1D $V_O$-mediated heterojunction of Cu-doped $TbFeO_3$ (CTFO) nanorods (1D), where $Cu^{2+}$ substitution at Fe-sites induces strain-mediated oxygen vacancies ($V_O$) for enhanced $CO_2$ chemisorption and electron trapping, and 0D $InVO_4$ (IVO) nanoparticles stacked onto CTFO to form a S-scheme junction, leveraging visible absorption of IVO and weak $CO^*$ adsorption for rapid desorption. This architecture couples $V_O$ with dimensional charge-transport advantages in which 1D nanorods enable axial electron flow, while 0D nanoparticles maximize surface sites and generate a built-in electric field at the interface. A built-in electric field in defect-engineered 0D/1D junctions enhances charge migration, optimizes charge-transfer paths, improves charge-separation efficiency, maintains a strong redox potential, and enhances photocatalytic performance. DFT total density of states (TDOS) in the IVO/CTFO study confirms the shift in the Fermi-level towards the conduction band, decreasing its bandgap, and resulting in enhanced light-capturing ability. The DFT free energy was calculated from $^*CO_2$ to $^*COOH$ in $CO_2$ photoreduction with IVO/CTFO, which has a lower energy barrier than CTFO, confirming the superior $CO_2$ conversion performance of IVO/CTFO. The investigation of the charge-transfer mechanism in the $V_O$-mediated 0D/1D junction is conducted using in situ DRIFTS, photoelectrochemical studies, and band-structure analysis. Notably, the recorded value of solar photo-reduction attained by this 0D/1D photocatalyst is the maximum reported so far. Thus, our work pioneers a defect-engineered dimensional heterojunction strategy with lattice-mismatched behavior to surpass selectivity and advance solar-driven $CO_2$ valorization without the incorporation of noble metals or sacrificial agents.

**Results and discussion**

The morphology of the IVO/CTFO 0D/1D junction was investigated employing a scanning electron microscope (SEM). The SEM image indicates that $InVO_4$ NPs are firmly attached to the surface of CTFO NRs, forming a well-integrated heterostructure, as shown in Figure 1(a). Additionally, Figures S1-S4 (Supplementary file) demonstrate the 1D and 0D morphology for TFO, CTFO, and IVO samples and the SEM-EDX plot of IVO/CTFO, respectively. Figure 1(b-d) indicates the TEM monograph with elemental mapping of IVO/CTFO, and a high-angle annular dark-field high-resolution scanning transmission electron microscopic (HAADF-STEM) image of IVO/CTFO, respectively. The HRTEM image displays that IVO NPs are stacked on CTFO NRs present in the IVO/CTFO binary junction, as depicted in Figure 1(e). This demonstrates the development of a precisely developed 0D/1D binary heterojunction, comprising 1D CTFO NRs embedded with 0D IVO NPs. The measured periodic spacings were 0.277 and 0.266 nm for (200) and (112) planes, corresponding to CTFO and IVO, respectively, as illustrated in Figure 1h.

The X-ray diffraction profile of the fabricated materials is depicted in Figure 1(f). Following XRD pattern, the sharp diffraction peaks for pristine TFO and Cu-doped CTFO were observed at 23.02°, 31.92°, 33.62°, 35.90°, 36.20°, 38.13°, 51.44°, 59.65°, which were assigned to the (101), (020), (210), (201), (201), (211), (231), and, (123) planes indicating that NRs are crystalline with *Pbnm* space group of perovskite TbFeO$_3$ (JCPDS card # 96-100-8092) attributed to orthorhombic perovskite structure [17]. The lattice parameters for TFO (a=5.60Å, b=7.63Å, c=5.32Å) and CTFO were determined. Further, it is noted that Cu substitution does not alter the crystallinity of TbFeO$_3$. Figure 2b indicates that the diffraction peak in CTFO is shifting towards lower angles at 31.92° and 51.44°, which is ascribed to the larger ionic radius of Cu$^{2+}$ than Fe$^{+3}$ (R$^{3+}_{Fe}$ = 0.64 Å, R$^{2+}_{Cu}$ = 0.73 Å), which caused an expansion in lattice parameters (a=5.72 Å, b= 7.76Å, c= 5.47Å) when Cu was substituted [18]. Moreover, the +2 valence state of Cu$^{2+}$ was retained and was not changed into +3 because the radius of the Cu$^{3+}$ ion (R$^{3+}_{Cu}$ = 0.54 Å) is smaller than that of R$^{3+}_{Fe}$ [19]. Similarly, diffraction peaks for IVO were generated at 28.27°, 31.04°, and 33.07° with (002), (200), and (112) planes, which are well indexed to pure orthorhombic (JCPDS card # 96-433-6636) without additional peak detection. The XRD pattern of the IVO/CTFO binary junction confirms resemblances to those of the IVO and CTFO, accompanied by prominent IVO and CTFO diffraction peaks. These profiles endorse the successful fabrication of IVO/CTFO V$_O$-mediated 0D/1D junction.

The intrinsic microstructural characteristics of TFO, CTFO, IVO, and IVO/CTFO binary junction were explored via the Raman scattering technique as depicted in Figure 1(g). It was noticed that numerous distinct scattering peaks at 255, 318, 328, 351, 436, 506, and 639 cm$^{-1}$ are present in TFO, which are assigned to the prominent vibration modes characteristic of the orthorhombic phase [20]. Moreover, Fe$^{+3}$ ions are Raman inactive due to their center of inversion behavior in the *Pbnm* configuration. The peaks at 318, 328, 436, and 506 cm$^{-1}$ are attributed to A$_g$ vibrations, while an intense peak at 227 and a small peak at 351 cm$^{-1}$ are credited to B$_{2g}$ and B$_{1g}$ vibrational patterns, respectively. The bands observed at 436 cm$^{-1}$ may correspond to the symmetric stretching, while the peak at 506 cm$^{-1}$ may arise due to symmetric bending vibration of the Fe-O bond. The pronounced broadness at the 639 cm$^{-1}$ region originates from a two-phonon process and might be associated with a crystal defect that is observed in polycrystalline samples. With Cu/Fe substitution in TFO, the Raman peaks became slightly broader, and the A$_g$ and B$_{2g}$ modes moved toward higher frequencies due to the Cu addition, which modified the local atomic structure in the TFO lattice. The lattice distortion caused by substituting Cu for Fe, with a lower mass, disrupted ionic equilibrium in TFO and led to partial ion deviation from equilibrium. Substitution of the Cu$^{2+}$/Fe$^{3+}$ ion may have induced V$_O$ to uphold charge neutrality [21]. Hence, V$_O$ may be expected to be the leading cause of the lattice distortion in TFO.

Additionally, IVO peaks at 251, 377, 424, and 918 cm$^{-1}$ in the InVO$_4$, corresponding to the characteristic vibration of the orthorhombic InVO$_4$ phase [22].

To confirm the presence of distinctive functional groups, the FTIR patterns of all the materials are depicted in Figure 1(h). The vibrational peak detected at ~3440 cm$^{-1}$ and in the 1620-1626 cm$^{-1}$ range is attributed to the O-H stretching of water vapors [23]. The peaks at 446 and 564 cm$^{-1}$ are ascribed to the vibrations of M-O and O-M-O (M = Tb, Fe, and Cu) [24]. The IVO peak at approximately 420 cm$^{-1}$ belongs to the In-O vibration, whereas the peak at 460 cm$^{-1}$ corresponds to the V-O-V band. Additionally, the prominent peak at 731 cm$^{-1}$ signifies the presence of the VO$_4^{3-}$ group [25]. Additionally, the doublet peaks at 900 and 950 cm$^{-1}$ are linked to the V-O-In and V-O vibrations, respectively [26]. Particularly, the coupling between CTFO and IVO decreases the intensity, which is associated with the suppression of specific vibration modes of the material, which are described with the penetration depth. As the penetration depth can vary in composites than the pristine material, this causes the inhibition in specific vibrational modes, which leads to a decreased in intensity [27].

Besides, zeta potential findings reveal that IVO possesses a negative surface charge; on the other hand, CTFO has a positive charge when dispersed in deionized water. This characteristic of coupling materials indicates that IVO underwent spontaneous deposition on CTFO owing to the difference in electrostatic attraction between them, as illustrated in Figure S5 [28].

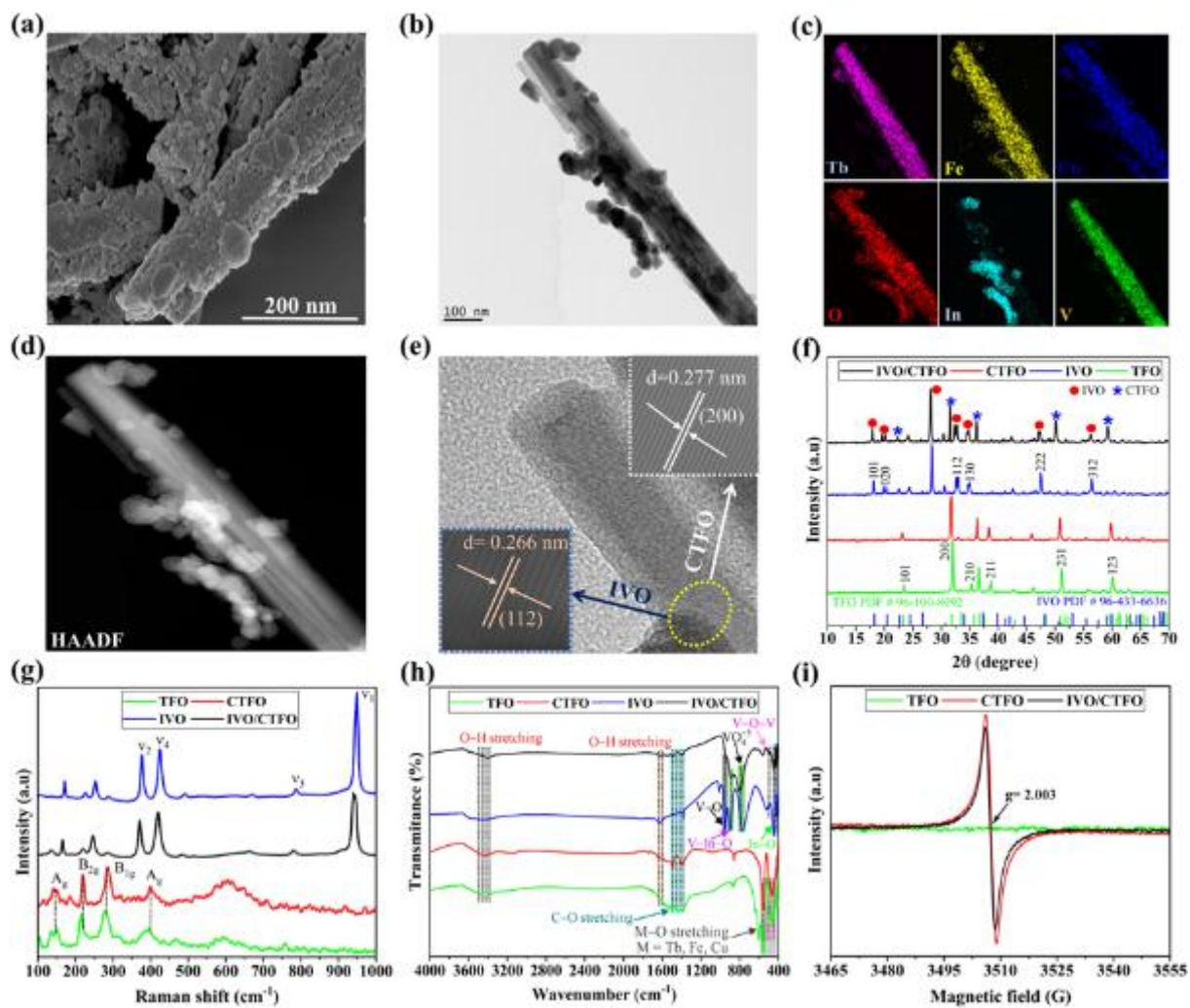

**Figure 1.** (a) SEM image, (b) TEM monograph, (c) TEM elemental mapping, (d) HAADF, and (e) HRTEM of IVO/CTFO, (f-h) XRD, Raman, and FTIR spectra of TFO, CTFO, IVO, and IVO/CTFO, respectively, and (i) EPR spectra of TFO, CTFO, and IVO/CTFO.

The creation of $V_O$ in photocatalysts is illustrated by the EPR analysis as demonstrated in Figure 1(i). EPR results depicted a strong signal at a g-value of 2.003 in CTFO and IVO/CTFO, while the EPR signal was not observed in pure TFO. Moreover, the EPR intensity of CTFO is stronger than IVO/CTFO, which indicates that Cu-doping causes the formation of $V_O$ in CTFO, which possesses more $V_O$ than the IVO/CTFO junction [29].

Spin-polarized total density of states (TDOS) calculations unveiled the profound electronic restructuring in the TFO-based architecture, as illustrated in Figure 2(a-c). In CTFO, extreme UP-spin dominance near the Fermi level inhibits carrier flow to a single spin channel, compelling charged species into spin-parallel configurations, which accelerate $e^-h^+$ pair recombination and cap the quantum efficiency. Thus, the 0D/1D junction (IVO/CTFO) was strategically abated through interfacial spin hybridization. DFT calculations revealed that the IVO/CTFO junction engineered the spin dynamics, fundamentally overcoming the photocatalytic limitations of spin-polarized CTFO by altering asymmetric carrier hauling into effective multichannel charge flow [30]. Furthermore, the DFT model of TDOS indicates, (a) the enhanced DOWN-spin intensity through Cu 3d-O 2p-V 3d orbital coupling, enabling dual-spin carrier contribution, (b) emergence of spin-moderated interface states within the bandgap, acting as recombination-suppressed pathways for fast electron transfer, and (c) shifting of the Fermi-level ($E_F$) towards the conduction band by reducing the bandgap, which leads to enhancing the potential of visible-light capturing [31].

Additionally, DFT calculations were carried out to elucidate the mechanisms underlying the migration of interfacial charges and the generation of the built-in electric field (BEF) at the IVO-CTFO interface. The electrostatic potential reveals a 1.02 eV interfacial work function increase in the IVO/CTFO junction (+7.48 eV) than the CTFO (+6.46 eV) and IVO (+6.56 eV) along the (112) and (200) planes, respectively, as shown in Figure 2(d-f). This increment stems from the three synergistic mechanisms: (a) Cu-induced charge localization generates surface dipoles by asymmetric electron density distribution, (b) $V_O$ creates a localized potential chamber that alters interfacial screening, and (c) IVO accumulation forms a perpetual charge-transfer dipole at the heterojunction. The subsequent potential gradient drives a directional electron flow toward CTFO, while positive charges are accumulated on IVO adjacent to the interface. The Φ value of IVO (112) is larger than that of CTF (200), while CTFO has a less positive $E_F$ than IVO, which facilitates electron migration from IVO to CTFO since it reaches an identical $E_F$ level at the interface [3b]. This electrostatic adjustment is computed by a 1.02 eV shift, which acts as a precise descriptor of the magnitude of interfacial charge redistribution. Besides, the work function directly correlates with higher interfacial electric field strength, as explained by the proportional enhancement in efficiency of charge separation [32].

The formation of an IVO/CTFO 0D/1D junction, as well as charge transfer dynamics, is depicted in Figure 2(g). Since IVO has a higher Fermi level than CTFO, electrons in IVO move toward CTFO until an equilibrium is reached. As a result, internal electric fields form, and band bending occurs at the interface. Upon light irradiation, photogenerated electrons in the CB of CTFO combine with the holes in the VB of IVO at the interface; meanwhile, spatially separated electrons on IVO and holes on CTFO participate in the photocatalytic reaction [33].

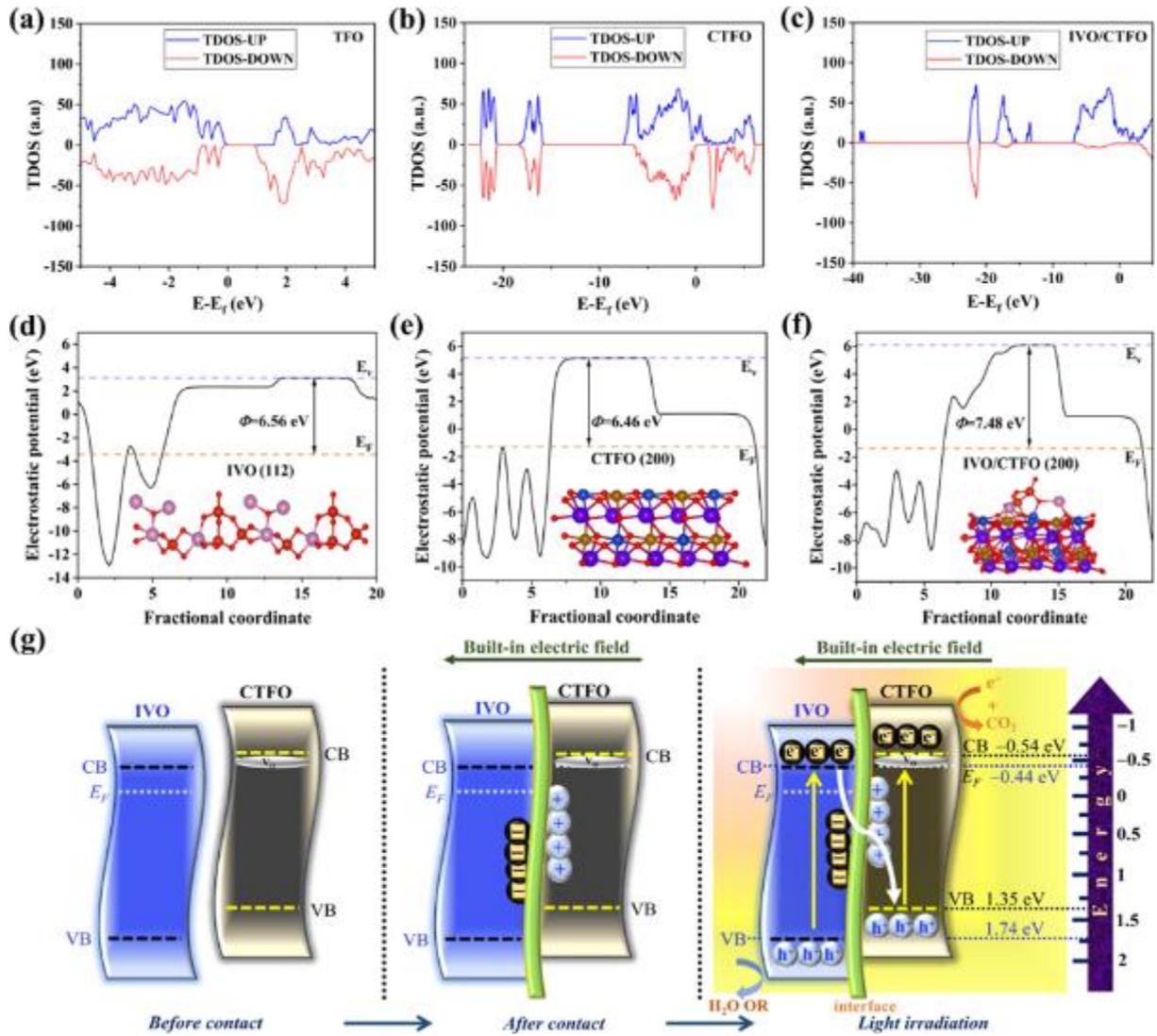

**Figure 2.** (a-c) total density of states of TFO, CTFO, and IVO/CTFO, respectively; (d-f) electrostatic potential of IVO, CTFO, and IVO/CTFO, and (g) Schematic diagram of the electron transfer mechanism between IVO and CTFO.

The charge density difference plots (Δρ) in CTFO exhibit asymmetric oscillations along the Z-axis, with sharp peaks of electron accumulation (Δρ > 0, up to +0.2 a.u) localized near Cu dopants and deep troughs of depletion (Δρ < 0, down to -0.3 a.u) at $V_O$-sites, as demonstrated in Figure 3(a, b). This reflects charge polarization, where Cu acts as an electron reservoir while vacancies trap holes. In the IVO/CTFO 0D/1D junction, a distinct interfacial Δρ shift occurs; electrons rapidly accumulate rapidly IVO (Δρ > 0, + 0.1-0.2 a.u) and decrease from CTFO (Δρ < 0, -0.1 to -0.3 a.u), generating a built-in electric field that drives electron transfer from IVO to CTFO. This built-in electric field is more effective at separating $e^-h^+$ pairs than in bulk materials. This directional charge separation, augmented by $V_O$ in CTFO acting as hole-trapping sites and Cu dopant enabling electron mobility, diminishes recombination and spatially isolates oxidation (IVO) and reduction (CTFO) sites, optimizing photocatalytic efficacy for redox reactions under visible light.

A schematic band structure demonstrates the sophisticated hierarchical engineering strategy within TFO NRs, where Cu doping (CTFO) lowers the conduction band minimum (CBM) to enhance electron affinity and reduction potential and concurrently stabilizes $V_O$, which serves as electron mediators and catalytic active sites for reactant activation, as demonstrated in Figure 3(c-e). The subsequent junction with IVO NPs (IVO/CTFO) creates a directional S-scheme junction, inducing a thermodynamic cascade. The photoexcited electrons transfer to the lower CBM of CTFO, whereas holes move to the higher valence band of IVO, making a spatial charge separation that drastically suppresses recombination of the $e^-h^+$ pair. Critically, the $V_O$ sustains at the interfaces, facilitating interfacial charge mediation and acting as multi-functional active sites, synergizing with the Cu-induced band modulation and heterojunction-driven carrier dynamics to optimize both oxidative (IVO) and reductive (CTFO) half-reactions. This synergetic effect of defect engineering ($V_O$), band alignment tuning (Cu-doping), and heterostructure design (IVO integrating) establishes a concerted electron-transfer pathway, transforming the composite into a highly efficient catalyst where vacancy-enabled kinetics, tailored redox potentials, and suppressed $e^-h^+$ pair recombination which collectively drive the efficacy beyond the sum of individual components, a paradigm for innovative catalytic material design.

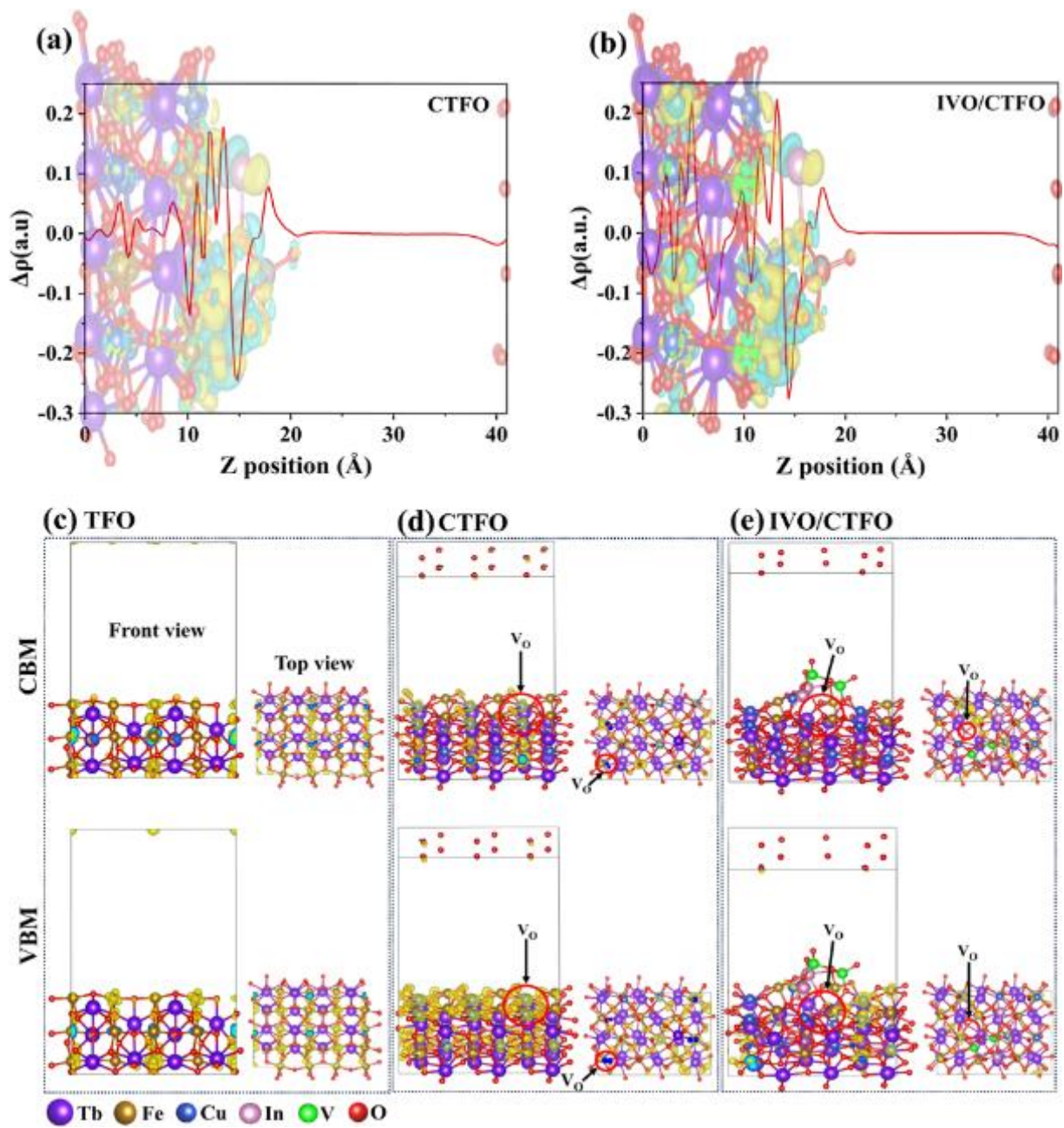

**Figure 3.** (a, b) Charge density difference plots with CTFO and IVO/CTFO, (c-e) electronic structure with VBM and CBM for TFO, CTFO, and IVO/CTFO

$N_2$-sorption was performed to analyze the specific surface area ($S_{BET}$) and pore size distribution of the material, as illustrated in Figure 4(a) and Table S1. Notably, the IVO/CTFO 0D/1D material displayed an enlarged specific surface area (SSA) compared to pure TFO, CTFO, and IVO samples. The observed difference may be associated with the generation of new pores caused by the stacking of IVO on the CTFO surface. The pore sizes of both CTFO and the IVO/CTFO are primarily found in the range of 10-20 nm, indicating that anchoring of IVO did not significantly alter the mesoporous structure of the CTFO NRs. Moreover, as demonstrated by Barrett Joyner Halenda (BJH), results are presented in Figure 4b(inset); the maximum pore size distribution increased following the incorporation of IVO on the CTFO surface. Because the reactants will have access to a larger effective surface area and additional transport routes, photocatalytic efficiency is expected to be enhanced. Correspondingly, the $CO_2$ adsorption capacity of IVO/CTFO was 3.96 cm$^3$/g, which is superior to that of CTFO and other pristine samples Figure 4(b). These results indicate that introducing IVO support significantly increases $S_{BET}$, pore volume, and $CO_2$ adsorption capacity.

Figure 4(c-f) presents the high-resolution deconvoluted XPS spectra for Tb 3d, Fe 2p, Cu 2p, and O 1s. Substituting Fe with Cu resulted in a slight increase in the bonding energy of the Tb 3d orbitals, shifting them to higher energy (to the left). In TFO, Tb, along with Fe, is present in the +3 valence state, verified by its corresponding binding energies. The XPS profile of Cu 2p for CTFO and IVO/CTFO depicts Cu $2p_{3/2}$ and Cu $2p_{1/2}$ peaks at ~932 and ~952 eV, respectively. that are attributed to $Cu^{+2}$ ions. The results validated that the Tb $3d_{5/2}$ and $3d_{3/2}$ peaks in the CTFO remained unaltered, demonstrating that, with $Cu^{+2}$ substitution, the oxidation state of Tb did not change [34]. The observed peaks at ~724.22 eV and ~711.12 eV correspond to the Fe $2p_{1/2}$ and $2p_{3/2}$ states, respectively, and are responsible for the spin-orbit splitting in TFO [35]. The asymmetry and breadth of Fe peak in CTFO denote that Fe exhibits various oxidation states ($Fe^{3+}$ and $Fe^{2+}$) [36]. The valence-state transition from $Fe^{3+}$ to $Fe^{2+}$ leads to the formation of $V_O$, which is essential for maintaining charge neutrality in CTFO [37]. Hence, XPS investigation confirms that substituting the Fe-site with an appropriate density of $Cu^{2+}$ ions can increase the density of defects, such as VO, thereby enhancing the photocatalytic efficiency of the IVO/CTFO junction [38]. Additionally, Cu has a lower electron density than Fe, thereby increasing the binding energy in CTFO and IVO/CTFO. Along with this, XPS peaks of In 3d and V 2p for IVO and IVO/CTFO are given in Figures S6, S7, respectively. Further, the O1s peaks at approximately 529.56, 530.14, and 530.78 eV for TFO, CTFO, and IVO/CTFO, respectively, and are attributed to oxygen species present in the lattice, indicating the existence of multiple Metal-O bonds as demonstrated in Figure 4(f). Thus, this binding energy is associated with the lattice oxygen that exists in the crystal [39], as $ABO_3$ perovskite usually comprises $V_O$ in the presence of species that are absorbed on the surface [18]. The peak between binding energies ~531.3 and 532.7 eV is correlated with the $V_O$ produced by $Cu^{2+}$ substitution in CTFO and IVO/CTFO. This $V_O$ peak demonstrates the efficacy of Cu substitution and results in the formation of surface $V_O$ [40]. This suggests that the addition of Cu at Fe sites plays a significant role in generating $V_O$, thus contributing to the enhanced photocatalytic reactions [21].

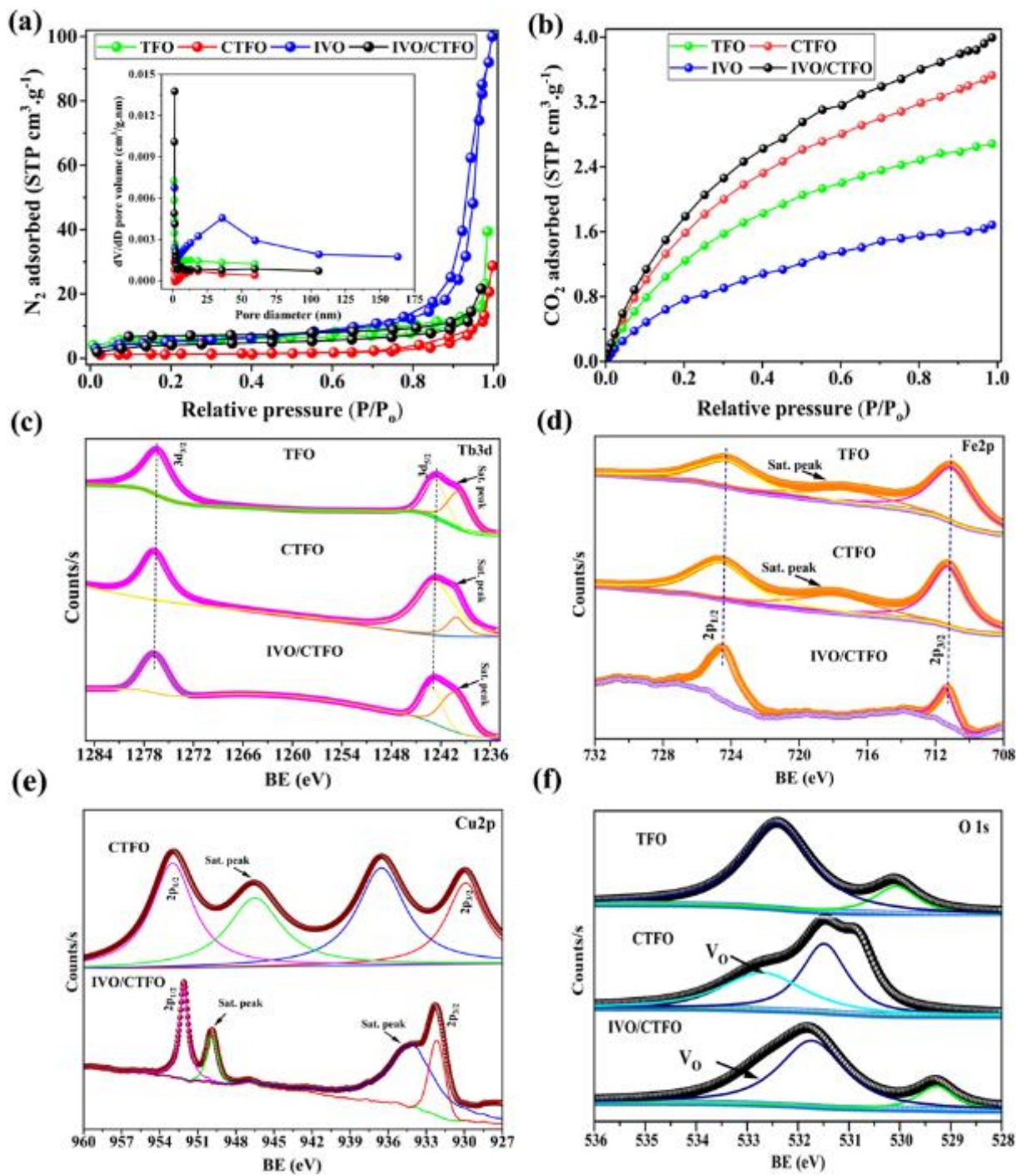

**Figure 4.** a) $N_2$ adsorption-desorption isotherms and pore distribution curves (inset), and b) $CO_2$ adsorption isotherms of TFO, CTFO, IVO, and IVO/CTFO, respectively, and deconvoluted XPS spectrum of (c) Tb3d, (d) Fe2p, (e) Cu2p, and (f) O1s.

Figure 5(a) demonstrates the energy band gap calculated by plotting photon energy (eV) Vs. $[F(R)h\nu]^{n/2}$ for all the samples. The band gaps of TFO and CTFO are 1.93 [8] and 1.89 eV, respectively, indicating that upon $Cu^{2+}$ addition, the band gap decreases to 1.89 eV, while IVO indicates a band gap of 2.18 eV [41]. $Cu^{2+}$ introduces new electronic states within the bandgap, acting as shallow acceptor levels near the valence band (VB). These states facilitate sub-bandgap transitions, allowing electrons to excite from the VB to Cu-induced states or from these states to the conduction band (CB) at lower energies than the intrinsic $Fe^{3+}$-$O^{2-}$ charge-transfer gap [42]. The ionic radius mismatch ($Cu^{2+}$: 0.73 Å vs. $Fe^{3+}$: 0.645 Å) distorts the $FeO_6$ octahedra, modifying Fe/Cu-O bond lengths and angles. This distortion reduces crystal field splitting, narrowing the bandgap. A lower charge on $Cu^{2+}$ may also promote $V_O$ for charge compensation, introducing defect states below the CB that further reduce the bandgap [43].

Photoluminescence (PL) analysis of TFO, CTFO, IVO, and IVO/CTFO was conducted to examine the recombination rate of the $e^-h^+$ pair. Figure 5(b) demonstrates the PL intensity spectrum, and the observed peak corresponds to electronic transitions from the conduction band to the valence band. The detected intensity difference analyzes the $e^-h^+$ pair recombination rate, by reflecting the order of intensity, IVO > TFO > CTFO > IVO/CTFO, which demonstrates that the 0D/1D junction possesses a superior $e^-h^+$ separation potential with the lowest recombination rate relating to IVO, TFO, and CTFO [35]. A substantial decrease in fluorescence intensity was observed for the IVO/CTFO junction, implying that rapid charge migration occurs between the hetero-interface (IVO/CTFO) and results in a decline of charge recombination. Furthermore, to validate this, a transient fluorescence time-resolved investigation was performed to determine the average decay time of charged carriers, as shown in Figure 5(c). An average lifetime of charge species is ordered as follows: IVO/CTFO (64.70 ns) > CTFO (54.81 ns) > TF (39.68 ns) > IVO (32.58 ns). This suggests that the extended decay time for IVO/CTFO leads to produce more photo-induced carriers to further increase in the $CO_2$ photoreduction reaction [9].

To rationalize the kinetics of charge separation and transfer, a systematic photo-electrochemical measurements were performed, as displayed in Figure 5(d-f). An amperometry (i-t) curve was performed to analyze the electron transfer process in both light-on and light-off modes. The intensity indicated a significant rise upon light illumination, and a sudden decline to lower levels once the light was switched off [44]. Figure 5(d) illustrates that the intensities of an amperometry curves for TFO and IVO are 0.57 µA and 0.38 µA, respectively. Conversely, the photocurrent in CTFO and IVO/CTFO was increased to 0.64 µA and 1.11 µA, respectively than TFO ad IVO, leading to slight recombination of photoproduced charged carriers. These results suggest that the addition of Cu and IVO stacking to CTFO enhances the $e^-h^+$ pair separation by forming an interface between IVO and CTFO. This interface leads to an improvement in the photocatalytic efficiency of the photocatalysts [45]. Additionally, a higher photocurrent response directs a faster migration of photogenerated carriers, as it is directly associated with the diffusion rate of electrons.

EIS tests are utilized to identify the interfacial charge migration. Figure 5(e) illustrates that the EIS semicircle radius of the IVO/CTFO junction is smaller than that of the other pristine samples, indicating that the IVO/CTFO junction has a lower interface charge transfer resistance and efficient $e^-h^+$ pair separation ability. The photoelectrochemical test results firmly establish superior photogenerated charge-separation and transfer properties in the IVO/CTFO 0D/1D junction compared to other samples.

The Mott-Schottky study was performed to ascertain the energy band structure for the photocatalysts. Figure 5(f) displays MS plot of the TFO, CTFO, IVO, and IVO/CTFO. The intersection of the slope with the x-axis may be clearly described as the conduction band edge (CBE) of the respective material. Nonetheless, the ascending trend observed in all the samples indicates, they are classified as n-type semiconducting materials [46]. By using Eq. (1) following a standard hydrogen electrode (SHE), the CBEs of TFO, CTFO, IVO, and IVO/CTFO were determined to be -0.56 eV, -0.63 eV, -0.52 eV, and -0.72 eV, respectively.

$$E_{NHE} = E_{Ag/AgCl} + E^o_{Ag/AgCl} - 0.059 \, (pH) \qquad (1)$$

Here, $E_{Ag/AgCl}$ denotes a potential measurement using Ag/AgCl as the reference electrode, with a standard potential (E°) of 0.198 at ambient conditions. The pH of a 0.5 M sodium sulphate ($Na_2SO_4$) electrolyte is approximately 6.8 [47]. Using Eq. (7), the Fermi level ($E_F$) was determined to be -0.37, -0.44, -0.34, and -0.53 V Vs NHE for TFO, CTFO, IVO, and IVO/CTFO, respectively. Subsequently, the CB of TFO, CTFO, IVO, and IVO/CTFO are calculated as -0.47, -0.54, -0.44, and -0.63, respectively, with 0.1 V deviations applied. Though their corresponding VBs are 1.46, 1.35, 1.74 eV for TFO, CTFO, and IVO, respectively, measured by their CB and $E_g$ through Eq. (2) [48]. These results indicate that the band edge potentials of both photocatalysts in the heterojunction, IVO and CTFO, are sufficient to generate the reactive oxygen species required for the photoreduction reaction.

$$E_{CB} = E_{VB} - E_g \qquad (2)$$

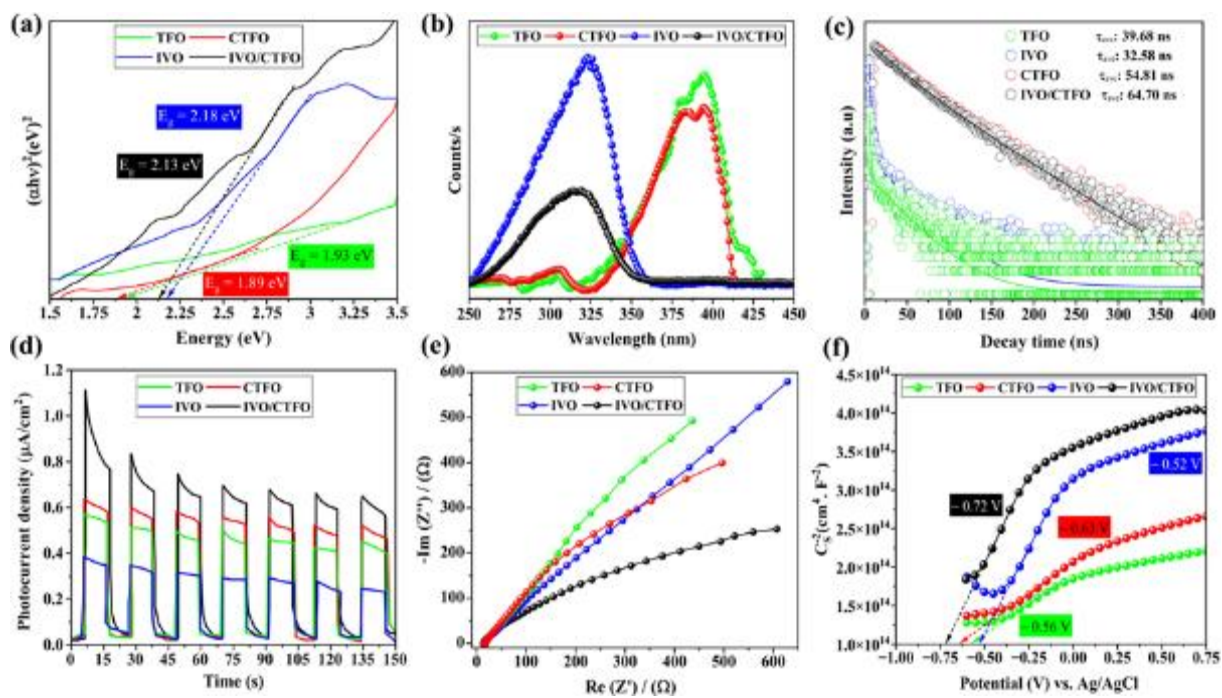

**Figure 5.** (a) Bandgap plot, (b) PL intensity, (c) TR-PL, (d) Photocurrent, (e) EIS, and (f) Mott-Schottky plot of TFO, CTFO, IVO, and IVO/CTFO

The photocatalytic $CO_2$ reduction performance was evaluated using the gas-solid mode. The gas was introduced into the gas chromatograph and quantified by comparing its peak intensity to that of a standard gas. For TFO, IVO, CTFO, and IVO/CTFO, the calculated CO rates were 23.61 and 18.89, 48.28 and 65.75 μmol g$^{-1}$ h$^{-1}$, respectively while $CH_4$ rates were, 2.73, 2.13, 3.94 and 2.67 μmol g$^{-1}$ h$^{-1}$, respectively as demonstrated in Figure 6 (a), where a 95.93% superior selectivity of CO was achieved with IVO/CTFO, Figure 6 (b). With IVO anchoring to the CTFO, a synergetic effect between VO and the lattice-mismatched behavior of NPs and NRs was observed, resulting in enhanced e$^-$h$^+$ pair separation and improved CO evolution rate and selectivity. The average CO evolution rate for IVO/CTFO was measured to be 1.36, 2.78, and 3.48 times higher than CTFO, TFO, and IVO, as depicted in Figure S8. Upon light excitation, the catalyst absorbs light energy, generating photogenerated electrons that jump to the conduction band. The adsorbed $CO_2$ on the catalyst surface is then reduced to CO with the assistance of protons and electrons via the reaction: $CO_2 + 2H^+ + 2e^- \rightarrow CO + H_2O$. The source of $H^+$ is through the oxidation of water.

To confirm the carbon source, a series of control experiments was conducted (Figure 6(c)). The results exhibited that no carbon-based products were detected in the absence of a photocatalyst, in the dark, under Ar gas, and without $H_2O$, endorsing that the generated CO originated from the photocatalytic reaction between $CO_2$, and $H_2O$ in presence of the catalyst. Hence, these results reveal that constructing the IVO/CTFO junction significantly increases the photocatalytic efficiency of $CO_2$ reduction. Additionally, to illustrate the superior efficiency of the IVO/CTFO 0D/1D junction, apparent quantum efficiency (AQE) for $CO_2$ photoreduction at wavelengths of 420, 450, and 475 nm was measured. Though. the AQE of IVO/CTFO was calculated to be 1.71% at 420 nm, which is consistent with the absorption spectrum shown in Figure 6(d). Furthermore, the stability of the catalyst is a crucial indicator for its catalytic performance. To study this, a recyclability trial for IVO/CTFO was conducted, as shown in Figure 6(e). The CO yields of IVO/CTFO showed no significant decrease after 6 cycles of irradiation with a 300W Xe lamp, indicating excellent photocatalytic activity and stability. As well, after the reusability experiment, structural and morphological analyses were performed to confirm the stability of IVO/CTFO. The catalyst was collected, washed, and characterized to assess any structural variations that may have occurred during the photoreduction experiments. The XRD profile, FTIR spectrum, and SEM image of IVO/CTFO, following the recycling tests, are presented in Figures S9 and S10, respectively. The XRD profile depicted in Figure S9 suggests that the crystal phase of IVO/CTFO remains unchanged before and after the $CO_2$ photoreduction reaction, confirming that the catalyst retains its crystallinity during the process, which justifies its structural stability. Moreover, the FTIR pattern is nearly identical; however, contact between $CO_2$ molecules and photocatalysts caused a slight peak shift at ~1400-1600 $cm^{-1}$, which endorses the structural stability of the IVO/CTFO (Figure S10). The SEM images (Figure S11) unveiled that the 0D/1D morphology of IVO/CTFO was preserved after the reaction, with no noticeable structural distortion. This suggests that the lattice structure and morphology of IVO/CTFO remained stable under reaction conditions. Though the evaluation of numerous features (BET surface area, PL lifetime, $CO_2$ adsorption, CO evolution rate, and selectivity) of IVO/CTFO for $CO_2$ photoreduction demonstrates its superiority over other materials, as demonstrated in Figure 6(f). The $^{13}CO_2$-labeling study confirmed that only $^{13}CO$ (m/z = 29) was observed by gas chromatography-mass spectrometry (GC-MS) when $^{13}CO_2$ was used as a carbon source (Figure 6(g)). Though similar outcomes are extensively reported in previous studies [49]. To highlight the advantages of IVO/CTFO in photocatalytic $CO_2$ reduction, a comparison was made with other reported gas-solid-phase $CO_2$ photoreduction materials, without the co-catalysts or sacrificial agents (Figure 6(h)). The IVO/CTFO displayed excellent performance for $CO_2$ photoreduction, highlighting its potential as an efficient and stable photocatalyst. Hence, the development of a $V_O$-mediated IVO/CTFO 0D/1D junction not only enhances its ability to reduce $CO_2$ but also provides remarkable strength and stability across repetitive experiments. This makes the $V_O$-mediated IVO/CTFO 0D/1D junction an excellent contender for real-world applications in the photoreduction of $CO_2$.

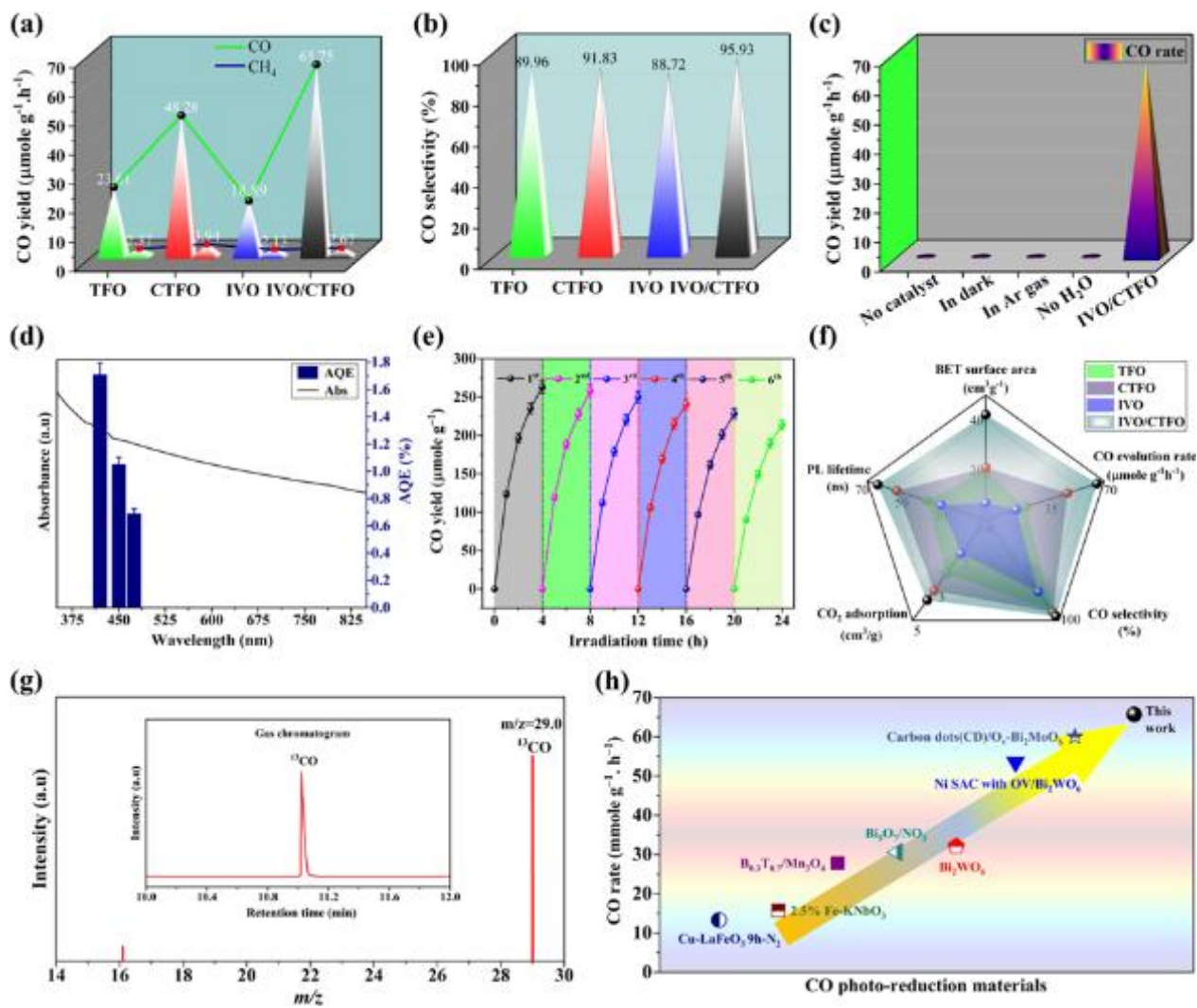

**Figure 6.** (a, b) CO evolution rate and selectivity, (c) series of control experiments, (d) apparent quantum efficiency, (e) reusability run for IVO/CTFO, (f) different evaluation parameters, (g) GC-MS spectrum of $^{13}CO$ evolution from $^{13}CO_2$ photo-reduction over IVO/CTFO, and (h) comparison of IVO/CTFO with other reported materials.

**Mechanism of $CO_2$ Photoreduction**

As the first step of $CO_2$ reduction, the adsorption and activation process is investigated. The temperature-programmed desorption of $CO_2$ ($CO_2$-TPD) was conducted to measure the adsorption potential of $CO_2$ on the catalyst surface, as shown in Figure 7(a). According to different desorption-temperature phases, adsorption are categorized into two types; physical adsorption and chemical adsorption during $CO_2$ activation [50]. The $CO_2$-TPD patterns of TFO, CTFO, and IVO/CTFO reveal distinct $CO_2$ adsorption behaviors critical for photocatalytic $CO_2$ reduction to CO. Pristine TFO exhibits weak $CO_2$ adsorption (100-300 °C), which can be assigned to physical adsorption with weak basic sites, while CTFO shows stronger binding (200-400 °C) due to Cu-induced $V_O$ enhancing $CO_2$ chemisorption. This analysis aligns with the BET results, as depicted in Figure 4(a), and can be attributed to the relatively low surface area of TFO and other samples. A bimodal desorption pattern was observed at the IVO/CTFO junction, characterized by weakly adsorbed $CO_2$ on the IVO surface (150-300 °C) and strongly bound species at the interface (350-500 °C), suggesting synergistic effects. Yet, this improved adsorption, coupled with enhanced separation of charge carriers at the IVO/CTFO junction, demonstrates the most efficient catalyst, as optimal $CO_2$ activation and interfacial electron transfer are crucial for efficient photoreduction of $CO_2$ to CO [32]. The findings underscore the significance of tailored surface modifications and heterojunction design in advancing $CO_2$ conversion technologies.

Deeper insights into the dynamic monitoring of adsorbed surface species and $CO_2$-derived intermediates in the photoreduction reaction were further investigated using in situ diffuse reflectance infrared Fourier transform spectroscopy (DRIFTS), as shown in Figure 7(b, c). This operando DRIFTS study reveals a critical mechanistic divergence in $CO_2$ reduction between CTFO and IVO/CTFO heterostructures: CTFO displays rapid accumulation of monodentate carbonate (b-$CO_3^{2-}$, 1350-1450 $cm^{-1}$) [51] and carboxylate (COOH, 1550-1650 $cm^{-1}$) within 10 minutes, demonstrating direct, however ineffective, $CO_2$ activation at Cu-vacancy sites that facilitates other small species formation and active-site blocking. Conversely, IVO/CTFO illustrates a kinetically orchestrated mechanism where IVO Bronsted-acidic V-OH groups first stabilize bicarbonate ($HCO_3^-$, 1400-1480 $cm^{-1}$) [29, 52] as a proton-regulated intermediate, delaying carboxylate appearance (20-30 min) while enabling concerted proton-electron migration that selectively drives $CO_2$ toward the carbonyl radical ($CO_2^-$, 1220-1280 $cm^{-1}$) intermediate. This radical stabilized at oxygen-rich interfaces at CTFO and underwent rapid desorption as CO due to IVO role as a proton-transfer modulator, which prevents carbonate poisoning by optimizing $H^+$ flux. Thus, IVO/CTFO 0D/1D junction achieves spatially decoupled catalysis: IVO governs proton management and initial activation, while CTFO specializes in electron transfer and CO desorption, suppressing parasitic pathways and achieving ~96% CO selectivity through interfacial synergy.

The free energy diagram, the Figure 7(d) reveals how the IVO/CTFO achieves an efficient thermodynamics for $CO_2$ to CO conversion by orchestrating a spontaneously downhill pathway: $CO_2$ activation initiates at an energy-neutral step (-0.89 eV), the free energy for the transition from *COOH to *CO for CTFO is 1.26 eV, which is higher than among all reaction steps from $CO_2$ to CO, thereby identifying it as the rate determining step [53]. Although, for IVO/CTFO, the rate determining step is consistent with CTFO, its free energy significantly reduces, followed by a uniquely stabilized *COOH (0.52 eV, significantly lower than conventional Cu catalysts due to interfacial $V_O$ acting as electron traps), which then cascades exothermally to *CO formation (-0.52 eV) without the typical kinetic barrier enabled by IVO Bronsted-acidic V-OH groups dynamically supplying protons for rapid dehydroxylation. Critically, the steep energy descent post CO (0.50 eV) signifies spontaneous product desorption and catalyst regeneration, circumventing the ubiquitous CO poisoning bottleneck; this arises from CTFO spin-polarized Cu sites weakening CO adsorption (via σ-repulsion from filled Cu 3d orbitals) while IVO electron injection maintains a low-energy vacancy-rich interface. The absence of a *CHO pathway energy trap further confirms exclusive CO selectivity, as the heterostructure bifunctional synergy IVO for proton-coupled electron transfer and CTFO for radical intermediate stabilization, eliminates parasitic routes, creating an electron-efficient thermodynamic slide from $CO_2$ to gaseous CO. Hence, the findings highlight the role of $V_O$ and IVO anchoring in accelerating dynamic charge transfer and the overall reduction process [54].

Based on the above results and conclusions, a proposed mechanism for charge transfer during efficient $CO_2$ photoreduction in the presence of the IVO/CTFO photocatalyst is illustrated in Figure 7(e). Visible light absorption generates an $e^-h^+$ pair in both CTFO (1.89 eV bandgap) and IVO (2.18 eV). Following an S-scheme band alignment, electrons move to the conduction band of CTFO, localized at $V_O$, serving as electron traps, while holes transfer to the valence band of IVO. Here, $V^{5+}/V^{4+}$ redox couples accelerate water oxidation, discharging protons ($H^+$) as depicted in the following equation: (3-7). Critically, lattice mismatch at the interface induces anisotropic strain, creating a built-in electric field that promotes charge separation and generates tensile-strained sites with undercoordinated metal cations ($Fe^{3+}$, $V^{5+}$). These strained sites adsorb and polarize $CO_2$ into metastable $CO_2^{•-}$ anions, while $V_O$ traps the electrons and weakens C=O bonds. Protons from water oxidation diffuse through the electrolyte and bind preferentially to protonated $V_O$ sites ($V_O$-$H^+$), forming Bronsted acid centers that facilitate sequential proton-coupled electron transfers. Strain geometry forces •COOH* intermediates into linear conformations, lowering the activation barrier for C-O bond cleavage. The morphology of NRs directs electrons along the [200] crystal axis, minimizing recombination, while the weak CO adsorption energy at $V_O$-sites ensures rapid desorption before over-reduction. This tripartite synergy, defect-mediated activation ($V_O$), strain engineering (mismatch), and morphology-driven transport (nanorods), kinetically steers the pathway toward CO with 95.93% selectivity, suppressing $H_2$ and $CH_4$.

**Step 1**: Formation of charged carriers ($e^-/h^+$)

$$\text{IVO/CTFO} + h\nu \rightarrow \text{IVO}(e_{CB}^-/h_{VB}^+)/\text{CTFO}(e_{CB}^-/h_{VB}^+) \qquad (3)$$

**Step 2:** Transfer of $e^-/h^+$ pair

$$\text{IVO}(e_{CB}^-/h_{VB}^+)/\text{CTFO}(e_{CB}^-/h_{VB}^+) \rightarrow \text{IVO}(e_{CB}^- + h_{VB}^+) + \text{CTFO}(e_{CB}^- + h_{VB}^+) \qquad (4)$$

**Step 3:** Oxidation-reduction reaction

$$\text{IVO}(h_{VB}^+) + H_2O \rightarrow O_2 + 2H^+ \qquad (5)$$

$$CO_2 + 2H^+ + 2e^- - \text{CTFO}(e_{CB}^-) \rightarrow CO + H_2O \qquad (6)$$

$$O_2 + 4H^+ + 2e^- \rightarrow 2H_2O \qquad (7)$$

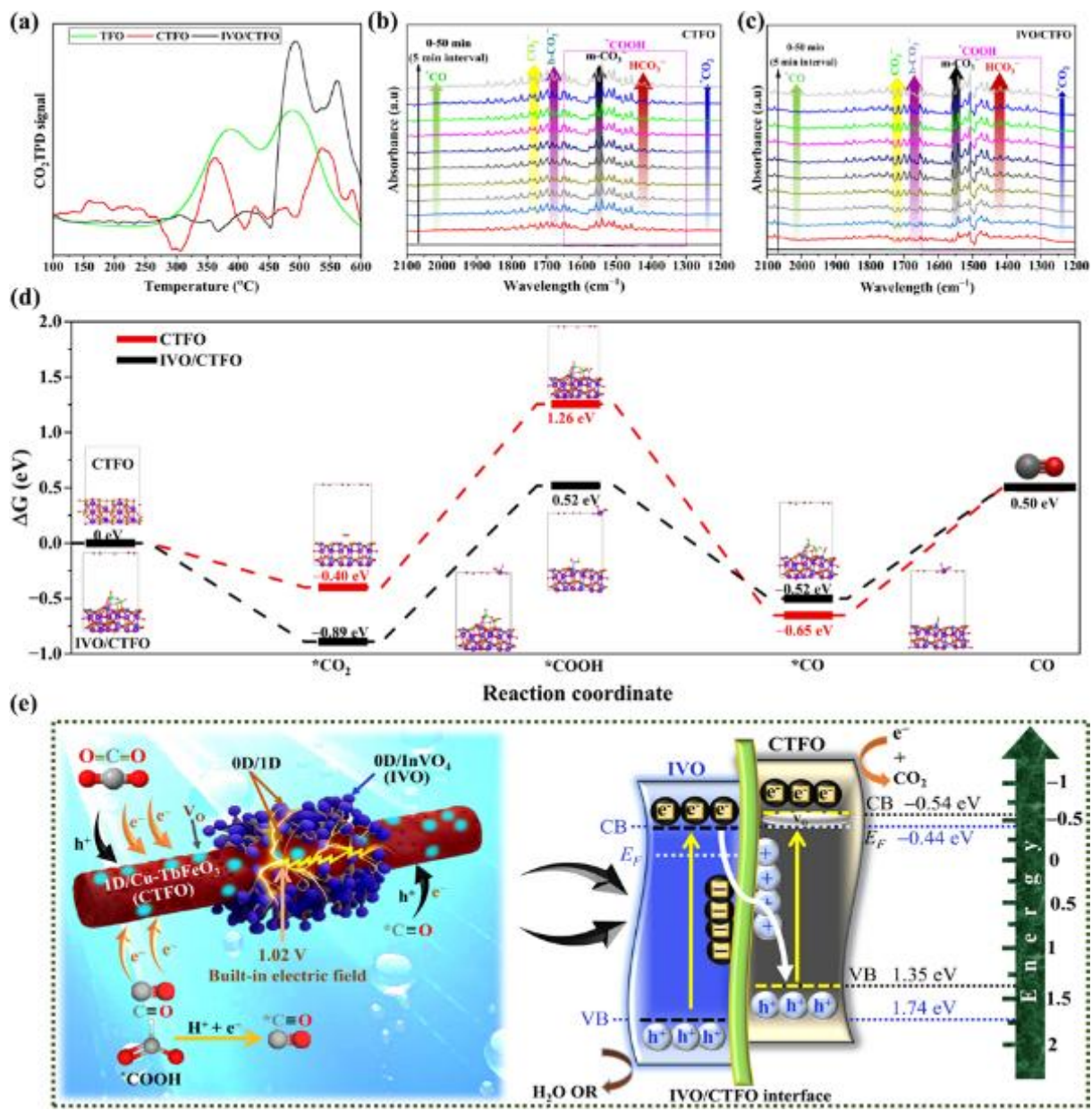

**Figure 7.** (a) CO$_2$-TPD, (b-c) DRIFTS spectra under simulated solar light illumination, (d) The calculated free energy diagram for the energy pathways from CO$_2$ photo-reduction to CO of CTFO and IVO/CTFO, and (e) photocatalytic CO$_2$ reduction mechanism over IVO/CTFO 0D/1D s-scheme junction

## Methods

**Materials:**

Terbium chloride hexahydrate (TbCl$_3$.6H$_2$O), iron chloride hexahydrate (FeCl$_3$.6H$_2$O), NaOH, copper chloride (CuCl$_2$), indium chloride (InCl$_3$), sodium vanadate (Na$_3$VO$_4$), ammonia solution, nitric acid, ethanol, and DI water. All chemicals were analytical grade and used as acquired.

**Synthesis of TbFeO$_3$ nanorods (NRs)**

The TbFeO$_3$ NRs were fabricated by the hydrothermal method, using ethylene glycol as a template and structural directing agent. In the first step of the synthesis process, a stoichiometric amount of TbCl$_3$.6H$_2$O (1 mmol) and FeCl$_3$.6H$_2$O (1 mmol) was homogeneously dissolved in a reaction mixture containing deionized water (45 mL), ethylene glycol (45 mL), and ethanol (10 mL). This homogeneous mixture was subjected to ultrasonication, followed by stirring for 6 hours. Ammonia solution was added gradually to reach a pH of 10. Then the dissolved sediments were washed repeatedly with an ethanol-water mixture. A few drops of 1 M NaOH were added to the mixture, which was then agitated for 1 hour to produce a stable, transparent solution. Following this, the attained reaction solution was shifted to a stainless-steel Teflon-lined autoclave reactor. Afterward, the solution was subjected to hydrothermal treatment for 72 hours at 180 °C. Subsequently, the fabricated product was decontaminated with an ethanol-water mixture and dried in a hot air oven overnight. To finish, the black product was annealed at 550 °C (ramp rate = 10 °C/min) for 3 hours to maintain its crystal structure. The powder was collected and labeled as TbFeO$_3$ NRs after cooling at ambient temperature.

**Synthesis of Cu-induced TbFeO$_3$ NRs**

The Cu-induced TbFeO$_3$ NRs were synthesized using the same process as for the preparation of TbFeO$_3$, with the stoichiometric addition of a CuCl$_2$ precursor.

**Synthesis of InVO$_4$ nanoparticles (NPs)**

0.4 g of sodium orthovanadate (Na$_3$VO$_4$) dodecahydrate was dissolved in 40 mL of DI water. By adding 0.586 g indium chloride (InCl$_4$) tetrahydrate to 20 mL of DI water, an aqueous solution was made. Then, this solution was gradually added to the Na$_3$VO$_4$ solution, yielding a yellow solution. With a 2 mol/L HNO$_3$ solution, the pH of the yellow solution was maintained at 2, and the mixture was stirred for 30 minutes. Consequently, it was transferred to a 100 mL stainless-steel Teflon-lined autoclave reactor and heated to 180 °C for 20 h. To remove impurities, InVO$_4$ nanoparticles were collected, rinsed repeatedly with ethanol and water, and dried at 80 °C overnight.

**Designing InVO$_4$@Cu-TbFeO$_3$ (IVO/CTFO) 0D/1D junction**

Among the different TbFe$_{1-x}$Cu$_x$O$_3$ (x=0.01, 0.03, 0.05, and 0.07) samples, the TbFe$_{1-0.05}$Cu$_{0.05}$O$_3$ (denoted as CTFO) NRs with the best CO$_2$ reduction performance were preferred as the host catalyst to be coupled with the InVO$_4$ (IVO) catalyst. Specifically, three solutions were made by adding 10%, 30%, and 50% w/v of InVO$_4$ into 50 mL of DI water. Each solution was poured into its respective stoichiometric solution of CTFO nanorods, maintained under continuous stirring. After stirring for 1 hour, the solution was heated to 150 °C and maintained at this temperature for 12 hours. The methods for washing and drying were identical to those described in the preceding section. Moreover, in this study, we report a 30% IVO rate with CTFO, which was superior to rates of 10% and 50%.

**Characterization of photocatalysts**

The microstructures and morphologies of the synthesized material were examined using a scanning electron microscope (German ZEISS Sigma 360) and a transmission electron microscope (TEM), FEI Talos F200X. The crystallinity of synthesized samples was assessed by X-ray diffractometer (XRD) (X'Pert PRO MPD, D8) with Cu-Kα radiation (λ=1.5406 Å) using a D/MAX-2000 at 40 kV and 30 mA. Brunauer-Emmett-Teller (BET) specific surface area and pore size distribution were evaluated using a MicrotracBEL (BELSORP-mini II). The $CO_2$ adsorption capacity was determined using $CO_2$-TPD on a BELSORP-MAX II (MicrotracBEL). X-ray photoelectron spectroscopy (XPS) (Thermo Scientific K-Alpha spectrometer with 300 W Al Kα radiation) was used to determine the chemical states of the elements present in the catalyst. Raman spectrum analysis (LabRam, HORIBA) was conducted using a laser line at λ = 532 nm in backscattering geometry at ambient conditions, with a 50× objective, over the range of 200-1000 $cm^{-1}$, with a 0.5 s integration time. The steady-state and transient fluorescence spectra (PL) were captured at excitation of λ = 325 nm using an Edinburgh FLS980 spectrofluorometer at room temperature. The UV-Vis diffuse-reflectance spectrum (UV-Vis DRS) was obtained for all materials using a Hitachi UH5700 spectrophotometer, while the Bruker Vertex 80V was employed for Fourier Transform Infrared (FTIR) analysis.

**$CO_2$ photoreduction**

The photocatalytic $CO_2$ reduction setup was designed for gas-solid-phase operation and conducted in a sealed quartz reactor at room temperature (25 ± 5 °C); notably, no sacrificial agent was employed. The AM 1.5 solar simulator served as a light source (MC-PF300C), positioned perpendicularly 2 cm above the quartz window of the reactor. Initially, 20 mg of catalyst was evenly distributed on the glass holder, which was then placed in the reactor and covered with a quartz window. Second, a high flow rate of 40 sccm of humidified $CO_2$ was purged into the chamber for 15 min, followed by a low flow rate of 5 sccm of humidified $CO_2$ before switching on the light source. During photocatalytic $CO_2$ reduction measurements, $CO_2$ gas was passed through a water bubbler to maintain a constant humidity level. After 4 hours of light illumination, the products were analyzed using a GC-7920 gas chromatograph (GC) equipped with a glass column packed with Porapak Q and a flame ionization detector (FID). The photocatalytic $CO_2$ photo-reduction stability test was conducted by repeating the identical protocol on the same sample every 4h for 6 cycles.

**Electrochemical Characterization**

Photoelectrochemical measurements were performed using a CHI760E electrochemical workstation (Shanghai Chenhua) with a standard three-electrode system. The working electrodes were prepared by using samples coated on ITO glass [55]. Pt-foil and Ag/AgCl (saturated KCl) electrodes were used as the counter and reference electrodes. The electrolyte was $Na_2SO_4$ solution (0.5M) and purged with $N_2$ gas for 1h before measurement. 5 mg photocatalyst was dispersed in 0.25 mL of ethanol, 1 mL of $H_2O$, and 10 μL of 5% Nafion (D-520) solution. Ultrasonic treatment was performed for 30 min. Finally, 80 μL of the above suspension was deposited on ITO with an active area of 1.5 cm$^2$ and dried at 60 °C. In the photocurrent measurements, a 300 W Xenon lamp with an AM 1.5G filter was used as the incident light, with a light intensity of 100 mW/cm$^2$ and no bias voltage applied. EIS tests were performed at the open-circuit voltage of each respective catalyst. Mott-Schottky (M-S) plots of photocatalysts were obtained in $N_2$-purged 0.5 M $Na_2SO_4$ electrolyte solutions using the same three-electrode system.

**Theoretical study**

All calculations with spin polarization were performed using the density functional theory, as implemented in the CP2K/Quickstep software package [56]. The core electrons were described with norm-conserving Goedecker-Teter-Hutter (GTH) [57] Pseudopotentials, and the valence electrons were described with Gaussian functions consisting of double-ζ polarized basis sets (m-DZVP) [58]. To improve the description of the on-site Coulomb interactions in CTFO, a Hubbard correction (DFT+U) with the effective U parameter of 3.5 eV was added to the Fe-d orbital. The van der Waals interactions were described using the Grimme DFT-D3 method with Becke-Johnson damping function [59]. The planewave cutoff for the finest real-space grid was set to 450 Ry, yielding total energies that converged to at least 0.001 eV/atom.

**Summary**

This work establishes a dual-defect-engineered IVO/CTFO heterojunction that achieves excellent $CO_2$ to CO photoreduction performance through synergistic lattice strain, oxygen vacancies ($V_O$), and dimensionality control. The 0D/1D architecture combines CTFO nanorods, where Cu doping creates $V_O$-sites for $CO_2$ adsorption and electron trapping, with IVO nanoparticles that facilitate rapid CO desorption. This design generates a built-in electric field at the lattice-mismatched interface, enhancing charge separation (64.70 ns carrier lifetime) and directional electron transfer, as evidenced by photoelectrochemical tests (1.11 μA photocurrent) and DFT models. The s-scheme heterojunction (IVO/CTFO) achieves 65.75 μmol g$^{-1}$ h$^{-1}$ CO production with 95.93% selectivity, outperforming existing noble-metal-free catalysts. Control experiments and isotopic labeling confirmed $CO_2$ as the carbon source. Stability over six cycles, as well as retained crystallinity (XRD) and morphology (SEM), underscores the structural stability. DFT free-energy calculations and in-situ DRIFTS elucidate the mechanism: $V_O$ lowers the *COOH formation barrier, while IVO's proton-transfer modulation suppresses byproducts. This defect-mediated dimensional heterojunction strategy paves the way for high-efficiency, selective solar fuel generation without the use of sacrificial agents.